\documentclass[conference]{IEEEtran}
\IEEEoverridecommandlockouts
\usepackage{cite}
\usepackage{amsmath,amssymb,amsfonts}
\usepackage{algorithmic}
\usepackage{graphicx}
\usepackage{textcomp}
\usepackage{xcolor,soul,framed}
\usepackage{comment}
\usepackage[linesnumbered,lined,boxed,commentsnumbered,ruled,longend, noend]{algorithm2e}

\SetCommentSty{mycommfont}
\SetKwInput{KwInput}{Input}                
\SetKwInput{KwOutput}{Output}              
\def\BibTeX{{\rm B\kern-.05em{\sc i\kern-.025em b}\kern-.08em
    T\kern-.1667em\lower.7ex\hbox{E}\kern-.125emX}}

\usepackage{hyperref}
\usepackage{fancyhdr,lipsum}

\fancypagestyle{mahmood}{%
   \fancyhf{} 
   
   \fancyhead[C]{Reprinting or republishing this material for the purpose of advertising or promotion, creating new collective works, reselling or redistributing to servers or lists, or using any copyrighted component in other works must adhere to IEEE policy. The paper has been accepted for publication at \textbf{MetaCom 2023}, and DOI will be provided as soon as possible.}
}%

\makeatletter
\let\ps@IEEEtitlepagestyle\ps@mahmood
\makeatother

\begin{document}

\title{Self-Sustaining Multiple Access with Continual Deep Reinforcement Learning for Dynamic Metaverse Applications
}

\author{
    \IEEEauthorblockN{
        Hamidreza Mazandarani, Masoud Shokrnezhad\textsuperscript{1}, Tarik Taleb\textsuperscript{1}, and Richard Li\textsuperscript{2}
    }
    \IEEEauthorblockA{
        \textit{\textsuperscript{1}Oulu University, Oulu, Finland; \textsuperscript{2}Futurewei Technologies, USA}\\
        hr.mazandarani@ieee.org; \{masoud.shokrnezhad, tarik.taleb\}@oulu.fi; richard.li@futurewei.com
    }
}

\maketitle

\begin{abstract}
The Metaverse is a new paradigm that aims to create a virtual environment consisting of numerous worlds, each of which will offer a different set of services. To deal with such a dynamic and complex scenario, considering the stringent quality of service requirements aimed at the 6th generation of communication systems (6G), one potential approach is to adopt self-sustaining strategies, which can be realized by employing Adaptive Artificial Intelligence (Adaptive AI) where models are continually re-trained with new data and conditions. One aspect of self-sustainability is the management of multiple access to the frequency spectrum. Although several innovative methods have been proposed to address this challenge, mostly using Deep Reinforcement Learning (DRL), the problem of adapting agents to a non-stationary environment has not yet been precisely addressed. This paper fills in the gap in the current literature by investigating the problem of multiple access in multi-channel environments to maximize the throughput of the intelligent agent when the number of active User Equipments (UEs) may fluctuate over time. To solve the problem, a Double Deep Q-Learning (DDQL) technique empowered by Continual Learning (CL) is proposed to overcome the non-stationary situation, while the environment is unknown. Numerical simulations demonstrate that, compared to other well-known methods, the CL-DDQL algorithm achieves significantly higher throughputs with a considerably shorter convergence time in highly dynamic scenarios.
\end{abstract}

\begin{IEEEkeywords}
Metaverse, 6G, Self-Sustainability, Non-Stationary, Multiple Access, Media Access Control (MAC), Adaptive AI, Continual Learning (CL), Deep Reinforcement Learning (DRL), Double Deep Q-Learning (DDQL).
\end{IEEEkeywords}

\section{Introduction}

The Metaverse is regarded as an advanced stage and the long-term vision of digital transformation that promises the creation of a 3-dimensional online virtual environment similar to the physical world \cite{tang_roadmap_2022}. This paradigm is expected to succeed the Internet in revolutionizing novel ecosystems of service provisioning in all walks of life (e.g., in extended reality, teleportation, unmanned mobility, and e-commerce), bringing even more challenges to the development of future wireless networks, which are already aimed at providing microsecond-level latency, bounded jitter, multi-gigabit-level throughput, extremely high reliability, and extremely low energy consumption \cite{giordani_toward_2020}. Given that the Metaverse environment will be comprised of a variety of worlds, each of which will provide different types of services, such quality standards need to be maintained in light of the fact that the Metaverse environment is constantly subject to change.

To face such highly dynamic environments where effective decisions must be made on a microsecond basis, various new paradigms have been introduced \cite{shokrnezhad2022near, shokrnezhad2023scalable}. As a potential strategy, one such paradigm is to employ mechanisms aiming to deliver "self-sustainability" as one of the driving factors toward the 6th generation of wireless communication systems (6G) \cite{alwis_survey_2021}. A self-sustaining network maintains its efficiency and effectiveness despite variable conditions. Unsurprisingly, a solution that fits well with the concept of self-sustaining networks is Adaptive Artificial Intelligence (Adaptive AI), where the mindset of \textit{once-in-a-lifetime train models} has been transformed into a new mindset in which models are \textit{continually re-trained} with new data and conditions. It is expected that Adaptive AI will be one of the most important enablers to facilitate the provision of emerging services, including Metaverse applications \cite{alwis_survey_2021}, and Gartner refers to it as one of the strategic technology trends in 2023 \cite{groombridge_gartner_nodate}.

Controlling multiple access to the frequency spectrum is one of the aspects of the self-sustaining feature that exists in 6G. In this scenario, a set of ever-fluctuating User Equipments (UEs) compete with one another for access to one or multiple frequency channels. Because these UEs are mobile and can be moved constantly from one access point to another at high speeds and frequencies, the number and type of them that have data to transmit over the frequency spectrum may vary over time. In addition, the traffic pattern might shift, either within a single UE from one moment to the next or across multiple UEs in terms of the active services seeking connection. The conditions of channels can change as well, influenced by a wide variety of noise sources and other environmental circumstances. Therefore, in order to realize future self-sustaining wireless networks, adaptive multiple access algorithms are essential.

In recent years, Deep Reinforcement Learning (DRL) has been leveraged for adaptive multiple access to the frequency spectrum. For instance, Yu \textit{et al.} \cite{yu_deep-reinforcement_2019} adopted DRL to design a Media Access Control (MAC) protocol without assuming the protocol of other coexisting UEs. They considered a heterogeneous environment with a slotted uplink channel. The same authors extended their work to non-uniform scenarios, in which channel sensing requires one time slot but information packet transmission requires multiple time slots \cite{yu_non-uniform_2021}. Jadoon \textit{et al.} \cite{jadoon_deep_2022} utilized DRL to optimize both throughput and packet age. Their research is compatible with machine-type communications on the assumption that the UEs are not saturated. Doshi \textit{et al.} \cite{doshi_deep_2021} formulated the coexistence of multiple base stations over a shared channel, optimizing the signal-to-interference-plus-noise ratio of UEs. Besides, Guo \textit{et al.} \cite{guo_multi-agent_2022} developed a solution for multi-agent scenarios to support delay-sensitive requests.

Although innovative techniques have thus far been proposed, the problem of adapting agents to a non-stationary environment has not been addressed. Since DRL cannot reuse previously learned knowledge, adapting to every change could be time-consuming, depending on the distance between context transitions. Therefore, the aforementioned approaches cannot be used in Metaverse scenarios considering their highly dynamic nature. This paper fills in the gap in the current literature by investigating the problem of multiple access in non-stationary, multi-channel, unknown environments in order to maximize the throughput of the intelligent agent by avoiding collisions with incumbent users. The non-stationarity is caused by intermittent changes in the set of active UEs. To solve the problem, a Double Deep Q-Learning (DDQL) technique empowered by Continual Learning (CL) is proposed, exploiting prior knowledge acquired throughout the agent's lifetime. Although a number of tools have been proposed to overcome non-stationary situations \cite{padakandla_survey_2021}, CL is the approach concerned with the adaptation of DRL-based agents \cite{khetarpal_towards_2022}.

The remainder of this paper is organized as follows: Section \ref{s_bck} introduces the background of DRL and CL. The system model and proposed approach are presented in Section \ref{s_app}. Finally, numerical results are illustrated and analyzed in Section \ref{s_sim}, followed by concluding remarks in Section \ref{s_con}.

\section{Background}\label{s_bck}

\subsection{Double Deep Q-Learning (DDQL)}
In Reinforcement Learning (RL), as a subset of machine learning techniques, an agent learns through trial and error how to optimize a given decision-making problem. The designer of the system specifies the reward function regarding the predefined design goals, and by learning and following the optimal strategy, the agent will maximize cumulative discounted rewards starting from any initial state. Q-Learning is probably the most recognized among the different algorithms introduced for model-free RL problems \cite{watkins_q-learning_1992}. Each state-action pair is assigned a numeric value in Q-Learning, known as the Q value, and this value is gradually updated by the following equation, which is the weighted average of the old value and the new information, that is
\begin{equation}\label{eq_bellman} 
Q(s_\tau, a_\tau)\;  \mathrel{{+}{=}} \; \sigma[Y_\tau^{QL} - Q(s_\tau, a_\tau)],
\end{equation}
where $s_\tau$ and $a_\tau$ are the agent's state and action at time slot $\tau$ respectively, $\sigma$ is a scalar step size, and $Y_\tau^{QL}$ is the target, defined by
\begin{equation}\label{eq_target}
Y_\tau^{QL}  = r_{\tau+1} + \gamma \; \text{max}_{a \in \boldsymbol{\mathcal{A}}} Q(s_{\tau+1}, a),
\end{equation}
where $r_{\tau+1}$ is the reward at time slot $\tau+1$, $\gamma \in [0,1]$ is a discount factor that balances the importance of immediate and future rewards, and $\boldsymbol{\mathcal{A}}$ is the set of actions.

\begin{figure}[t!]\centering
\includegraphics[width=3.3in]{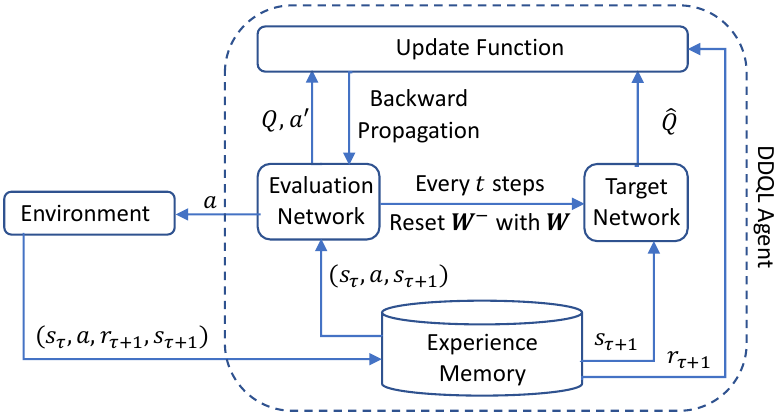}
  \caption{DDQL agent.}
  \label{fig_ddql}
\end{figure}

Since the majority of worthwhile problems are too large to discover all possible combinations of states and actions and learn all state-action values, Double Deep Q-Learning (DDQL) is a ground-breaking alternative to approximate them, wherein 1) Deep Neural Networks (DNNs) are used to approximate Q values, and 2) the selection and evaluation of actions are decoupled \cite{hasselt_deep_2016}. In DDQL, the state is provided as the input, and the $Q$ function of all possible actions, denoted by $Q(s, .; \boldsymbol{\mathcal{W}})$, is generated as the output, where $\boldsymbol{\mathcal{W}}$ is the set of DNN parameters. The target of DDQL is as follows:
\begin{equation}\label{eq_DDQL_target} 
Y_\tau^{DDQL}  = r_{\tau+1} + \gamma \; \widehat{Q}(s_{\tau+1}, a', \boldsymbol{\mathcal{W}}^-_\tau),
\end{equation}
and the update function of $\boldsymbol{\mathcal{W}}$ is
\begin{equation}\label{eq_DQL_bellman} 
\footnotesize
\boldsymbol{\mathcal{W}}_{\tau+1} = \boldsymbol{\mathcal{W}}_{\tau} + \sigma[Y_\tau^{DQL} - Q(s_\tau, a_\tau; \boldsymbol{\mathcal{W}}_\tau)]\nabla_{\boldsymbol{\mathcal{W}}_\tau} \cdot Q(s_\tau, a_\tau; \boldsymbol{\mathcal{W}}_\tau),
\end{equation}
where $a' = \text{argmax}_{a \in \boldsymbol{\mathcal{A}}} Q(s_{\tau+1}, a, \boldsymbol{\mathcal{W}}_\tau)$. In this model, $\boldsymbol{\mathcal{W}}$ represents the set of weights for the main (or evaluation) $Q$ and is updated in each step, whereas $\boldsymbol{\mathcal{W}}^-$ is for the target $\widehat{Q}$ and is replaced with the weights of the main network every $t$ steps. In other words, $\widehat{Q}$ remains a periodic copy of $Q$. The DDQL agent is represented in Fig. \ref{fig_ddql}. To improve the efficiency, the observed transitions are stored in a memory bank known as the experience memory, and the neural network is updated by randomly sampling from this pool. 

\subsection{Continual Learning (CL)}
In real-world settings, especially in the ever-changing Metaverse ecosystem, it is anticipated that the probability transition function or reward function will change over the lifetime of the agent. This non-stationarity necessitates a distinction between training and testing periods. Recent advances in DRL have demonstrated impressive efficiency in a variety of tasks, but they frequently pivot around an agent that focuses on mastering a narrow task of interest. Besides, after any significant change, RL agents frequently require additional training to adapt to the new environment, and even after this training, they lack the ability to generalize to new variations, even for simple problems. Therefore, dynamic environments necessitate novel learning mechanisms distinct from other types of learning (such as meta or multi-task learning). CL is concerned with the adaptation of the RL agent to the evolution of these environments over time \cite{khetarpal_towards_2022}. In CL, the system contains $\mathcal{M}$ contexts (or tasks) $\mathcal{T}_m$ sequentially, where $m \in \{1,2,...,\mathcal{M}\}$. 

While \textit{catastrophic forgetting} (i.e., losing performance on old tasks after learning new tasks) is a critical issue that CL seeks to address, \textit{interference} is another issue that has yet to be handled. Interference occurs when two tasks have incompatible (or even contradictory) optimal actions for the same observation. To effectively manage these challenges, Kessler \textit{et al.} \cite{kessler_same_2022} proposed an algorithm, named OWL. This algorithm 1) employs a single network with a shared feature extractor but multiple heads, parameterized by linear layers to fit individual tasks; and 2) flushes the experience replay buffer prior to beginning learning for a new task. At the time of testing, task selection is approached as a multi-armed bandit problem in order to adaptively choose the optimal policy. Additionally, the authors employed the Elastic Weight Consolidation (EWC) mechanism to prevent forgetting between tasks. This algorithm slows down learning on specific weights based on their significance to previously observed tasks. The OWL algorithm is the foundation of our method, which is described in the following section.

\section{Proposed Approach}\label{s_app}
\subsection{System Model}

\begin{figure}[!t]
\centerline{\includegraphics[width=2.5in]{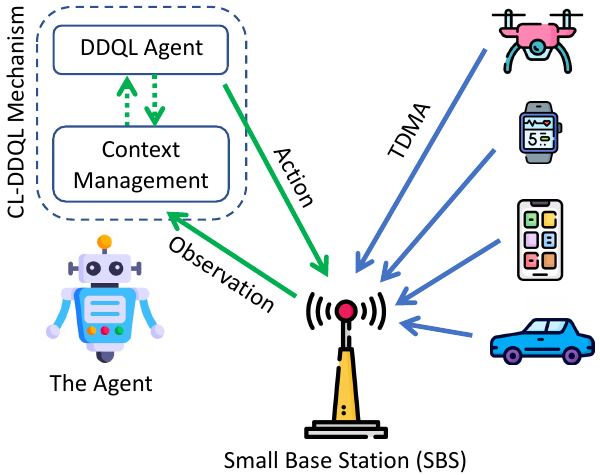}}
\caption{System model.}
\label{fig_sm}
\end{figure}

We consider a single small cell covered by a Small Base Station (SBS) with $n \in \{0, …, \mathcal{N}\}$ User Equipments (UEs) competing over ${C}$ time-slotted channels (see Fig. \ref{fig_sm}). Except for one (i.e., the CL-DDQL agent, or simply \textit{the agent}), all UEs periodically transmit their packets using the Time-Division Multiple Access (TDMA) protocol. For example, a headset may transmit visual recordings to its control center every second. The environment is non-stationary due to the fluctuating number of active TDMA users. Changes in the number of active Metaverse users can be attributed to a variety of factors, including users' mobility and bandwidth-saving strategies. In the headset example, if the user is inactive, data may be transmitted every 10 seconds. A context (or task) is defined as a collection of active UEs with unique identifiers on specific channels (e.g., UE $0$ on channel $1$ and UE $1$ on channel $2$ would constitute a simple context). Consequently, context transitions occur when a UE enters or leaves a channel. It is assumed that the agent is informed of the arrivals and departures of other UEs via SBS. However, the agent is unaware of the transmission profiles, so it must learn to coexist with these UEs.

The agent's transmissions are independent of SBS to avoid unnecessary signaling overhead in scheduling grant decoding. However, it relies on the SBS's ACK signals issued at the end of each packet transmission (or channel sensing) to indicate successful transmission (or channel idleness). The transmission of control messages is assumed to occur over a separate, collision-free channel. Similar to Yu \textit{et al.} \cite{yu_non-uniform_2021}, we assume UEs with variable-length packets, where $k \in \{1, ..., \mathcal{K}\}$ represents the packet length, as this is more practical than fixed-size packets \cite{yu_deep-reinforcement_2019}. Unlike Yu \textit{et al.} \cite{yu_non-uniform_2021}, however, our approach takes multiple channels into account, making it even more applicable in high-bandwidth Metaverse environments. To coexist successfully with other UEs, the objective function of the agent is to maximize its throughput by utilizing idle time slots in the channels.

\subsection{Agent Customization}
The first step in exploiting an RL agent for a particular problem is to define the agent's action, reward, and state space.  We define the action space as set $\mathcal{A} = \{a: (k, c) | k \in \{1, ..., \mathcal{K}\}, c \in \{1, ..., \mathcal{C}\}\}$, where $a: (0, c)$ points to sensing channel $c$ for one time-slot, and $a: (k > 0, c)$ denotes the transmission of a packet with length $k$ on channel $c$. Since the agent is designed to maximize its throughput, the reward is equal to the length of successfully transmitted packets. In the case of sensing channel $c$, the observation set would be $\boldsymbol{O}$ = \{\textit{Busy, Idle}\}, whereas it would be $\boldsymbol{O}$ = \{\textit{Success, Collision}\} in the case of packet transmission. The state of the agent is the sequence of the most recent $\mathcal{H}$ (observation, packet length, channel) tuples. To further enhance the Q function, we employ the dueling mechanism in the DDQL agent's evaluation network (Fig. \ref{fig_ddql}). Two estimators are utilized in this mechanism: one for the state value function and one for the state-dependent action advantage function. The primary advantage is the ability to generalize learning across actions without modifying the learning algorithm, which improves policy evaluation in the presence of numerous actions with similar values.

\begin{figure}[t!]\centering
\includegraphics[width=3in]{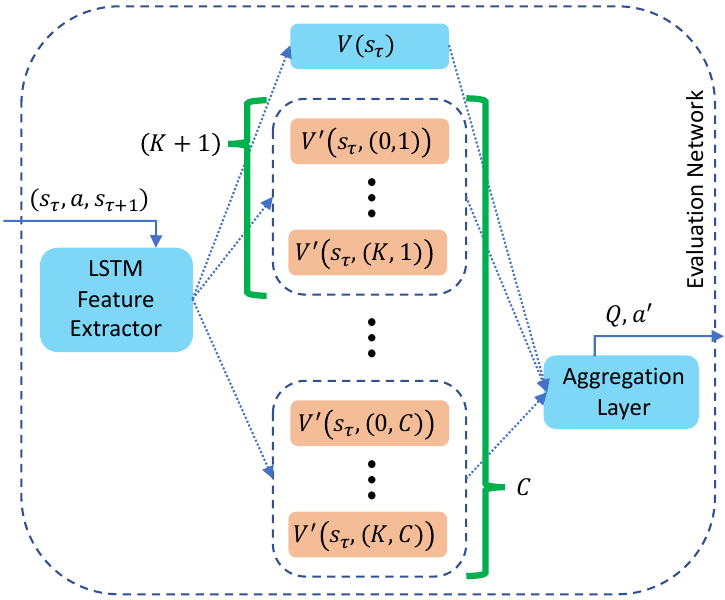}
\caption{Evaluation network of DDQL (Fig. \ref{fig_ddql})}
\label{fig_evl_net}
\end{figure}

The evaluation network mechanism of the DDQL agent is detailed in Fig. \ref{fig_evl_net}. In this module, the state is fed to a Long Short-Term Memory (LSTM) feature extractor in order to discover patterns that are consistent across all contexts. Afterwards, two sequences (or streams) of fully interconnected layers are utilized. The streams are designed to provide separate estimates of the state value function and the state-dependent action advantage function, denoted $\mathcal{V}$ and $\mathcal{V'}$, respectively. The two streams are combined to produce Q values as the final step. Additionally, to update the Q function, the target function in \eqref{eq_DDQL_target} must be transformed due to the non-uniformity of action lengths:
\begin{equation}\label{eq_customized_target} 
Y_\tau^{\star}  = \frac{(1 - \gamma ^ {d_{\tau}})}{(1 - \gamma) \; {d_{\tau}}}  \;  r_{\tau+1} +  \gamma ^ {d_{\tau}}  \; \widehat{Q}(s_{\tau+1}, a', \boldsymbol{\mathcal{W}}^-_\tau),
\end{equation}
where $d_\tau$ is the length of the action. For actions of length one (sensing the channel or sending a single time slot packet), \eqref{eq_DDQL_target} and \eqref{eq_customized_target} are obviously equivalent. However, future time slots are discounted for larger packages.

\subsection{CL Mechanism}
To accommodate the non-stationary nature of the environment, the proposed DDQL agent should be modified to remember previously learned contexts and rerun the training procedure for new contexts. In order to accomplish this, a CL mechanism is proposed and detailed in Algorithm. \ref{alg_clddql}. In this algorithm, $\mathcal{T}$ represents the lifetime of the agent, whereas $\epsilon'$ and $\widetilde{\epsilon}$ are small positive integers used to control the $\epsilon$-greedy mechanism. Through each step, if the agent is informed of a new context ($\phi$) by SBS, it saves the current experience memory and weights before examining the recorded contexts ($\boldsymbol{\Omega}$). If $\phi$ has been viewed previously, its experience memory and weights are loaded again. Otherwise, these parameters and $\epsilon$ reset after step 1. Following this, the reward and observation are collected and used to update the weights via the experience memory. Note that the action in each iteration is chosen by the $\epsilon$-greedy policy that follows the evaluation function of the corresponding agent with probability $(1-\epsilon)$ and chooses a random action with probability $\epsilon$. During the training process, the probability decreases linearly from $\epsilon$ to $\widetilde{\epsilon}$.

\begin{algorithm}[t!]\label{alg_clddql}
\caption{CL-DDQL}
\KwInput{$\mathcal{T}$, $\epsilon'$, and $\widetilde{\epsilon}$}
$\boldsymbol{\Omega} \leftarrow \emptyset$, $\boldsymbol{\mathcal{W}} \leftarrow \mathbf{0}$, $\boldsymbol{\mathcal{W}^-} \leftarrow \mathbf{0}$, $\epsilon \gets 1$, $memory \gets \{\}$ \\
\For{each $\tau$ in $[0:\mathcal{T}]$}
{
    \If{new context $\phi$ is announced}
    {
        save the current context memory and weights \\
        \If{$\phi \notin \boldsymbol{\Omega}$}
        {
            $\boldsymbol{\Omega} \leftarrow \boldsymbol{\Omega} \cup \{\phi\}$\\
            reset $\boldsymbol{\mathcal{W}}, \boldsymbol{\mathcal{W}^-}, memory$, and $\epsilon$
        }
        \ElseIf{$\phi \in \boldsymbol{\Omega}$}
        {
            reload $\boldsymbol{\mathcal{W}}, \boldsymbol{\mathcal{W}^-}$, and $memory$ of $\phi$
        }
    }
    $\zeta \gets$ generate a random number from $[0:1]$ \\
    \If{$\zeta > \epsilon$}
    {
        $(k, c) \gets argmax_{a \in \boldsymbol{\mathcal{A}}} Q(s_{\tau}, a, \boldsymbol{\mathcal{W}})$
    }
    \Else
    {
        select a random $(k, c)$ from $\boldsymbol{\mathcal{A}}$
    }
    transmit the packet, and get $O_\tau$ and $r_{\tau+1}$ \\
    calulate $s_{\tau+1}$ \\
    $memory \gets memory \cup \{(s_{\tau}, (k,c), r_{\tau+1}, s_{\tau+1})\}$ \\
    choose a sample form $memory$, and train the agent \\
    \If{$\epsilon > \widetilde{\epsilon}$}
    {
        $\epsilon \gets \epsilon - \epsilon'$
    }
}
\end{algorithm}

\section{Evaluation}\label{s_sim}

Within this section, a numerical analysis into the effectiveness of the proposed CL-DDQL method is conducted. The hyper-parameters and configurations are listed in Table \ref{tab_sim_par}. In order to test the efficacy of our strategy, we carried out a series of experiments on a computer running a 64-bit operating system that was equipped with 16 NVIDIA Tesla V100 Graphics Processing Units (GPUs) and 10 gigabytes of Non-Volatile Memory express (NVMe) storage. PyTorch was utilized to effectively implement both the evaluation and target networks. In each experiment, comparisons are made between the CL-DDQL, DDQL, and Random algorithms. The only difference between DDQL and CL-DDQL is that the CL-DDQL agent has a context management mechanism, whereas the DDQL algorithm lacks remembrance, so each announced context appears to be new to it. Finally, the Random agent transmits a packet that has a random length over a random channel. This will be accomplished without any prior knowledge or any specific adjustments being made to the configuration.

To compare algorithms, we use three metrics: normalized agent throughput, collision rate, and convergence time. The normalized agent throughput is computed by summing the length of the packets successfully transmitted over the last 1000 time slots (excluding headers) and dividing it by the maximum achievable throughput sum within the same window. The collision rate is the ratio of collision observations to total observations in the last 1000 time slots. Time between the occurrence of a context change and when the agent's throughput reaches a steady state is the convergence time. All metrics are averaged over 10 simulation rounds. In the first scenario, we establish fixed context transition points and fixed context specifications in order to better illustrate the efficacy of our strategy. Then, in the second scenario, we evaluate our scheme in a more realistic setting by assuming stochastic transition points and context specifications.

\begin{table}[t!]
\caption{Training Configuration.}
\begin{center}
\begin{tabular}{|c|c|}
\hline
\textbf{Parameter} & \textbf{Value} \\
\hline
Maximum packet length ($\mathcal{K}$) & $10$ time slots \\
Packet header size & $0.5$ time slot \\
State size ($\mathcal{H}$) & $20$ experiences \\
Capacity of experience memory & $1000$ experiences \\
Batch size & $32$ \\
Learning rate & $0.001$ \\
Exploration parameters $\widetilde{\epsilon}$, $\epsilon'$ & 0, 0.005 \\
Approximator model & \begin{tabular}{@{}c@{}}LSTM with $64$ units + \\ fully connected with $32$ units \end{tabular}  \\
Training frequency & \begin{tabular}{@{}c@{}} Each time slot \end{tabular}  \\
\begin{tabular}{@{}c@{}} Target network update frequency \\ \end{tabular} & Every $20$ steps \\
\hline
\end{tabular}
\label{tab_sim_par}
\end{center}
\end{table}

\subsection{Scenario 1: Fixed Change Points}
\begin{figure*}[!t]
\centerline{\includegraphics[width=7in]{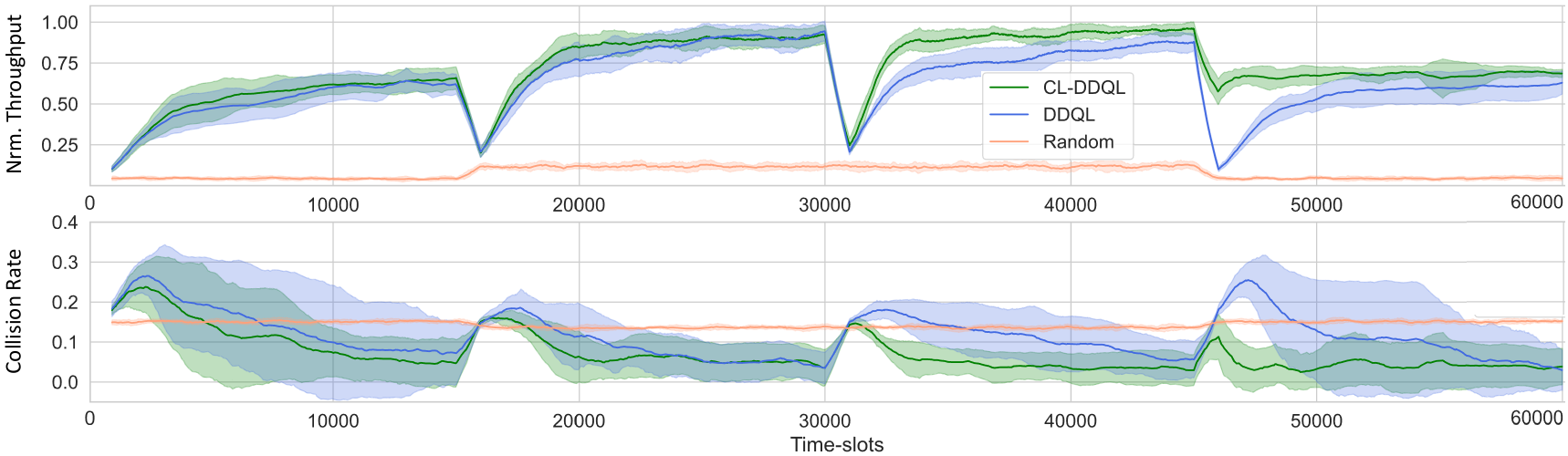}}
\caption{Normalized throughput and collision rate vs. time slots for CL-DDQL, DDQL, and Random}
\label{fig_snr01}
\end{figure*}

In this scenario, it is assumed that context transitions occur at specific times, as outlined in Table \ref{tab_scr01}. ($k, \tau, f, c$) identifies a TDMA UE that transmits a packet of size $k$ beginning on the $\tau$-th time slot of each frame of size $f$ on channel $c$. Clearly, the first and final quarters of the simulation take place in the same context; therefore, the CL-DDQL agent should utilize its prior knowledge of the first context when encountering it again. Fig. \ref{fig_snr01} verifies that the CL-DDQL agent possesses the required backward transfer capability for non-stationary environments. In addition, the agent utilizes its forward transfer capability when confronted with novel contexts. Despite the fact that the second and third contexts are distinct from the first, the pre-trained feature extractor enables the CL-DDQL algorithm to converge significantly more quickly than the conventional DDQL algorithm. In addition, the figures reveal that DDQL has greater variations in all metrics, which is highly undesirable in wireless networks. Evidently, Random, the method with the lowest complexity, is also inefficient.

\begin{table}[t!]
\caption{Scenario 1: UE Profiles.}
\begin{center}
\begin{tabular}{|c|c|c|}
\hline
\textbf{UE ID} & \textbf{Profile ($k, \tau, f, c$)} & \textbf{Period} \\
\hline
1 & (3, 0, 8, 0) & [0: $\mathcal{T}$/4] and [3$\mathcal{T}$/4: $\mathcal{T}$]  \\
2 & (4, 3, 8, 1) & [0: $\mathcal{T}$/4] and [3$\mathcal{T}$/4: $\mathcal{T}$]  \\
3 & (4, 0, 9, 0) & [$\mathcal{T}$/4: $\mathcal{T}$/2]  \\
4 & (2, 4, 9, 1) & [$\mathcal{T}$/4: $\mathcal{T}$/2]  \\
5 & (4, 0, 9, 1) & [$\mathcal{T}$/2: 3$\mathcal{T}$/4]  \\
6 & (2, 4, 9, 0) & [$\mathcal{T}$/2: 3$\mathcal{T}$/4]  \\
\hline
\end{tabular}
\label{tab_scr01}
\end{center}
\end{table}

\subsection{Scenario 2: Stochastic Change Points}

In this scenario, context shifts occur intermittently. When a UE arrives on a channel, it remains active at a rate of $1 / \beta$ according to an exponential distribution. After its departure, a new UE will replace it. The new UE has a novel (i.e., previously unseen) profile with probability $P$ and a repetitive profile with probability $1-P$. In our simulations, we set $P$ to 0.5. Moreover, the parameters of UE profiles are sampled from a set of distributions, namely $\{\mathcal{U}_{1,4}, \mathcal{U}_{4,8}, \mathcal{U}_{8,12}, \mathcal{U}_{1,\mathcal{C}}\}$ respectively, where $\mathcal{U}$ represents the Uniform distribution. Two experiments are defined by hyper-parameters $C$ (number of channels) and $\beta$ (mean duration of UE existence in the network). In the first experiment, the number of channels is set to 2, but $\beta$ varies from 20 to 100 percent of simulation time (and so the duration of contexts varies). In the second experiment, $\beta$ remains constant while the number of channels ranges from 1 to 5.

\begin{figure}[!t]
\centerline{\includegraphics[width=3in]{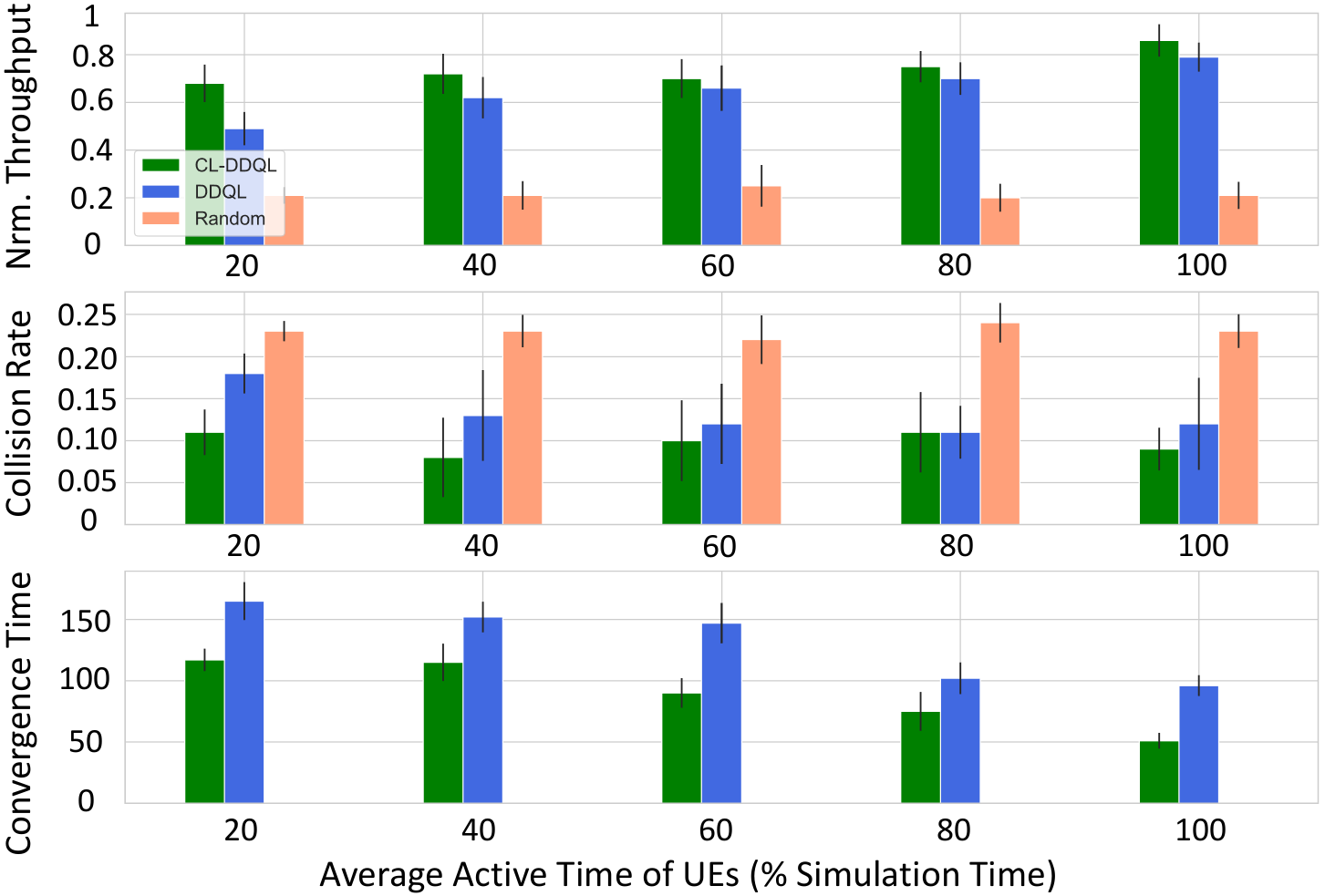}}
\caption{Normalized throughput, collision rate, and convergence time vs. the average active time of UEs for CL-DDQL, DDQL, and Random. The results are the all-time average of the values.}
\label{fig_scr02_exp01}
\end{figure}

\begin{figure}[!t]
\centerline{\includegraphics[width=3in]{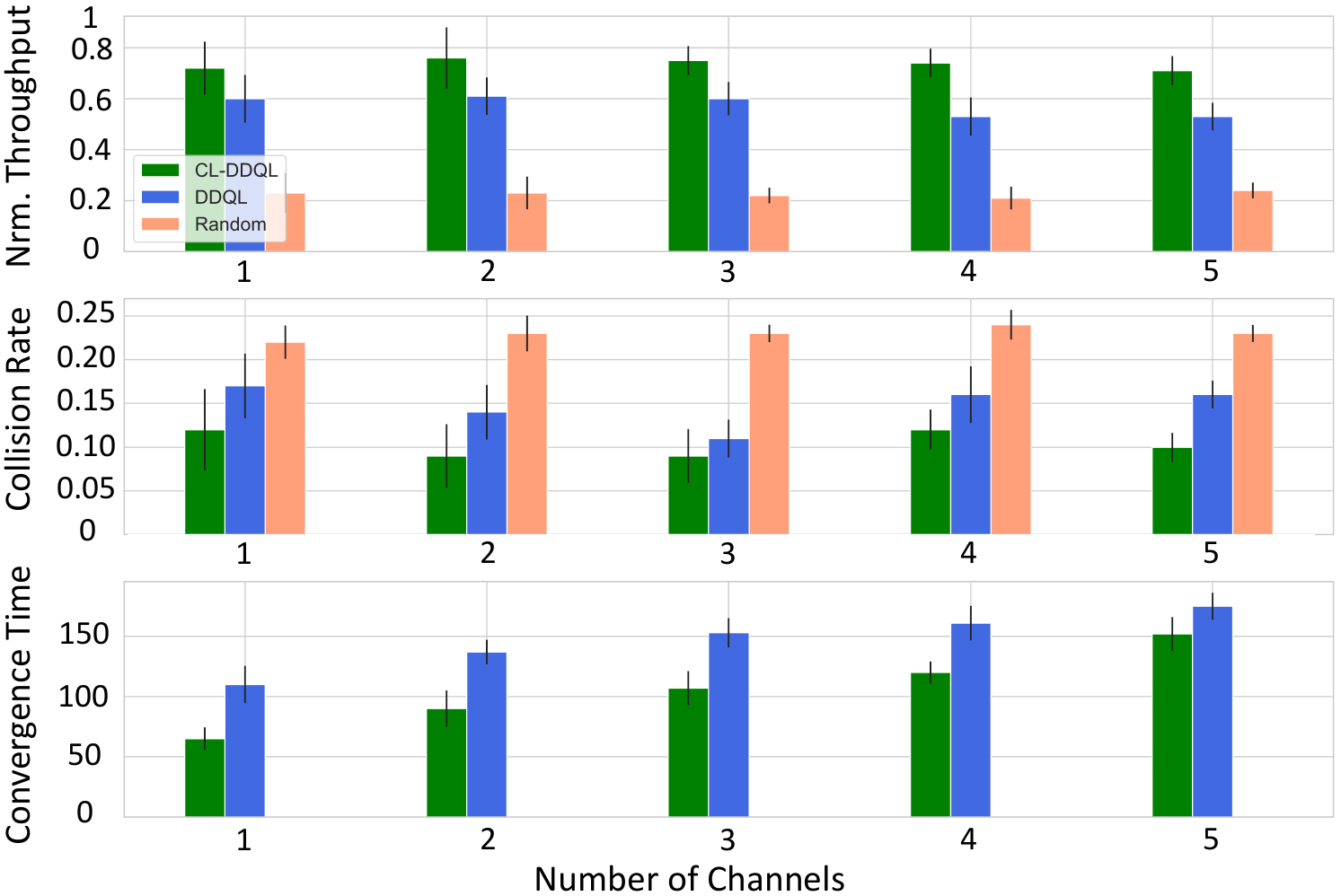}}
\caption{Normalized throughput, collision rate, and convergence time vs. the number of channels for CL-DDQL, DDQL, and Random. The results are the all-time average of the values.}
\label{fig_scr02_exp02}
\end{figure}

As Fig. \ref{fig_scr02_exp01} demonstrates, the more frequent the context transitions (lower values for $\beta$), the more continual learning improves the performance. This is due to the increased likelihood of encountering repetitive contexts. In addition, the performance of CL-DDQL is hardly impacted by an increase in the rate at which contexts are transited, making it suitable for the highly dynamic environments of the Metaverse. Nonetheless, both algorithms perform better in environments with less variability. For the second experiment, Fig. \ref{fig_scr02_exp02} illustrates that as the number of channels increases, the CL-DDQL algorithm becomes marginally more advantageous than DDQL. Notwithstanding, the performance of the two algorithms is not significantly impacted by the number of channels, leading us to conclude that while a greater number of channels provides more idle time slots for the agent, it also increases problem dimensions and thus the number of novel contexts to be explored.

\section{Conclusion}\label{s_con}
In this paper, the multi-channel multiple access problem was investigated while taking into account a non-stationary scenario in which the number of active UEs might shift over the course of time. The primary objective was to achieve maximum throughput while avoiding collisions with existing users. Initially, we introduced DRL and CL as two Adaptive AI mechanisms that could aid in the realization of self-sustaining networks. Afterward, a DDQL-based agent that is empowered by CL is designed. This agent is in charge of making decisions regarding spectrum access, such as adjusting a channel and modifying the length of the packet that needs to be transmitted. The effectiveness of the suggested agent was proved by the numerical results. Compared to other well-known methods, the CL-DDQL algorithm was shown to achieve significantly higher throughputs with a considerably shorter convergence time in highly dynamic unknown environments.

As a potential future work, we intend to tackle the problem by incorporating non-stationary channels with varying state probability distribution functions. In addition, we plan to enhance the CL-enabled DDQL-based method for accessing the spectrum for semantically-aware scenarios in which transmitting a subset of active UEs is sufficient to construct the parallel near-real-world experience, which could be a game-changer for bringing the Metaverse into existence by filtering out redundant data and maximizing the utilization of scarce communication resources.

\section*{Acknowledgment}
This research work is partially supported by the European Unions Horizon 2020 Research and Innovation Program through the Charity project under Grant No. 101016509, the Academy of Finland 6G Flagship program under Grant No. 346208, and the Academy of Finland IDEA-MILL project under Grant No. 352428.

\bibliographystyle{IEEEtran}
\bibliography{IEEEabrv,main}

\end{document}